\documentclass[]{aastex62}
\graphicspath{{./}{figures/}}

\received{May 9, 2018}
\revised{June 11, 2018}
\accepted{\today}
\submitjournal{ApJ}

\shorttitle{Average opacity calculation for CCSNe}
\shortauthors{Nagy}
\begin{document}

\title{Average opacity calculation for core-collapse supernovae}

\correspondingauthor{Andrea P. Nagy}
\email{nagyandi@titan.physx.u-szeged.hu}

\author[0000-0002-0786-7307]{Andrea P. Nagy}
\affiliation{Department of Optics and Quantum Electronics, University of Szeged\\
Dom ter 9, Szeged, 6720, Hungary}
\affiliation{Konkoly Observatory of the Hungarian Academy of Sciences\\
Konkoly-Thege ut 15-17, Budapest, 1121, Hungary}

\begin{abstract}

Supernovae (SNe) are among the most intensely studied objects of modern astrophysics, but due to their complex physical nature, theoretical models are essential to understand better these exploding stars, as well as the properties of the variation of the emitted radiation. One possibility for modeling SNe light curves is the construction of a simplified semi-analytic model, which can be used for getting order-of magnitude estimates of the SN properties. One of the strongest simplification in most of these light curve models is the assumption of the constant Thomson-scattering opacity that can be determined as the average opacity of the ejecta.
Here we present a systematic analysis for estimating the average opacity in different types of core-collapse supernovae (CCSNe) that can be used as the constant opacity of the ejecta in simplified semi-analytic models.
To use these average opacities self-consistently during light curve (LC) fit we estimate their values from hydrodynamic simulations. In this analysis we first generate MESA \citep{paxton11,paxton,paxton15,paxton18} stellar models with different physical parameters (initial mass, metallicity, rotation), which determine the mass-loss history of the model star. Then we synthesize SN LCs from these models with the SNEC hydrodynamic code \citep{morozova} and calculate the Rosseland mean opacity in every mass element. Finally, we compute the average opacities by integrating these Rosseland mean opacities. As a result we find that the average opacities from our calculations show adequate agreement with the opacities generally used in previous studies.

\end{abstract}


\section{Introduction} \label{sec:intro}
Core-collapse supernovae form a diverse group of supernova explosion events, but all of them are believed to originate from the death of massive stars ($M > 8 M_{\odot}$). The classification of these events is based on both their spectral features and light-curve properties \citep{class,turatto,prentice}. Core-collapse SNe are divided into several groups: Type Ib/Ic, Type IIP, Type IIb, Type IIL, and Type IIn.

The various types of CCSNe are thought to determine the explosion of the star with different progenitor properties like radii and ejected mass. However the mass loss could be the key parameter \citep{heger}, which define the final stage of the progenitor: stars having higher initial mass are able to lose their H-rich envelope, leading to Type IIb or Type Ib/Ic events, unlike the lower mass stars that create Type IIP explosions \citep[e.g.][]{sukh}.   

One possibility to estimate the initial pro\-perties of these events is to fit their LC with semi-analytic models that contain many assumptions (e.g.: homologous expansion, spherically symmetric ejecta, constant density profile). These calculations are able to produce a wide variety of SN light curves depending on the choice of the initial parameters, such as the ejected mass ($M_{ej}$), the initial radius of the progenitor ($R_0$), the kinetic energy ($E_{kin}$), and the mass of the synthesized $^{56}$Ni ($M_{Ni}$), which directly determines the emitted flux at later phases. These order-of-magnitude estimates for the basic parameters may also be useful to narrow the parameter regime in more detailed hydrodynamical simulations.

One of the strongest simplification in most of these LC models is the constant opa\-city approximation, which assumes that the opacity of the ejecta is constant in both space and time and also equals to the Thompson-scattering opacity ($\kappa_{Th}$). The advantage of this approximation is that $\kappa_{Th}$ depends only on the average chemical composition of the supernova ejecta. But it should keep in mind that the final che\-mical composition of the progenitor can be influenced by se\-ve\-ral physical processes, which determine the mass loss history or the convective mixing of the exploding star. In spite of this fact, in the literature the generally used approach is that the average opacity is only a model parameter that has no strong connection with the chemical composition. Thus, for example, its value should be about 0.3 - 0.4 cm$^2$/g for a H-dominated Type IIP SN, approximately 0.2 - 0.3 cm$^2$/g for H-poor Type IIb/Ib explosions and roughly 0.06 - 0.1 for He-poor Type Ic events \citep[e.g.][]{chev,taddia}.  
  
In this paper we aim to calculate the average opacities that are self-consistent with the chemical composition of typical core-collapse SNe. Because the proper choice of the opacity could be relevant in reducing the uncertainty of some fitting parameters, such as $M_{ej}$ and $R_0$ that are strongly correlated with the assumed Thompson-scattering opacity \citep[e.g.][]{nagy1}. 
  
This paper is organized as follows. In Sect. 2 and 3 we briefly describe the applied method, and the estimated average opacities, respectively. In Sect. 4 we present a verification for our results. Finally, Sect. 5 summarizes the main conclusions of this paper.

\section{Calculating the average opacity} \label{sec:opac}
During this work we approximate the ave\-ra\-ge opacities via synthesized light curve mo\-dels. The internal structure of the progenitor stars are derived from stellar evolution mo\-dels created by the MESA code \citep{paxton11,paxton,paxton15,paxton18} with different initial physical parameters from pre-main sequence up to core-collapse. It should keep in mind that the opacity calculation in this phase is based on the combination of opacity tables from OPAL, \cite{ferguson}, and \cite{cassisi}.

The subsequent hydrodynamic evolutions are followed by the 1D Lagrangian supernova explosion code, SNEC \citep{morozova}. This code solves the hydrodynamic and diffusion radiation transport in the expanding envelopes of the core-collapse supernovae. In SNEC the recombination processes and the heating due to radioactive decay of $^{56}$Ni and $^{56}$Co are also taken into account. In all calculated models the “thermal bomb” explosion scheme are used, in which the total energy of the explosion is injected into the model with an exponential decline both in time and mass coordinate. The SNEC code calculates the opacity in each grid point of the model from Rosseland mean opacity tables for different chemical compositions, temperatures, and densities. For this process an opacity minimum is also needed for the code. In our simulations the opacity boundary was 0.24 cm$^2$/g for the pure metal layer and 0.01 cm$^2$/g for the solar composition envelope \citep{bersten}. Thus, in SNEC the opacity at each time and grid point is chosen as the maximum value between the calculated Rosseland mean opacity and the opacity minimum for the corresponding composition.

\begin{figure}[!ht]
\centering
\plotone{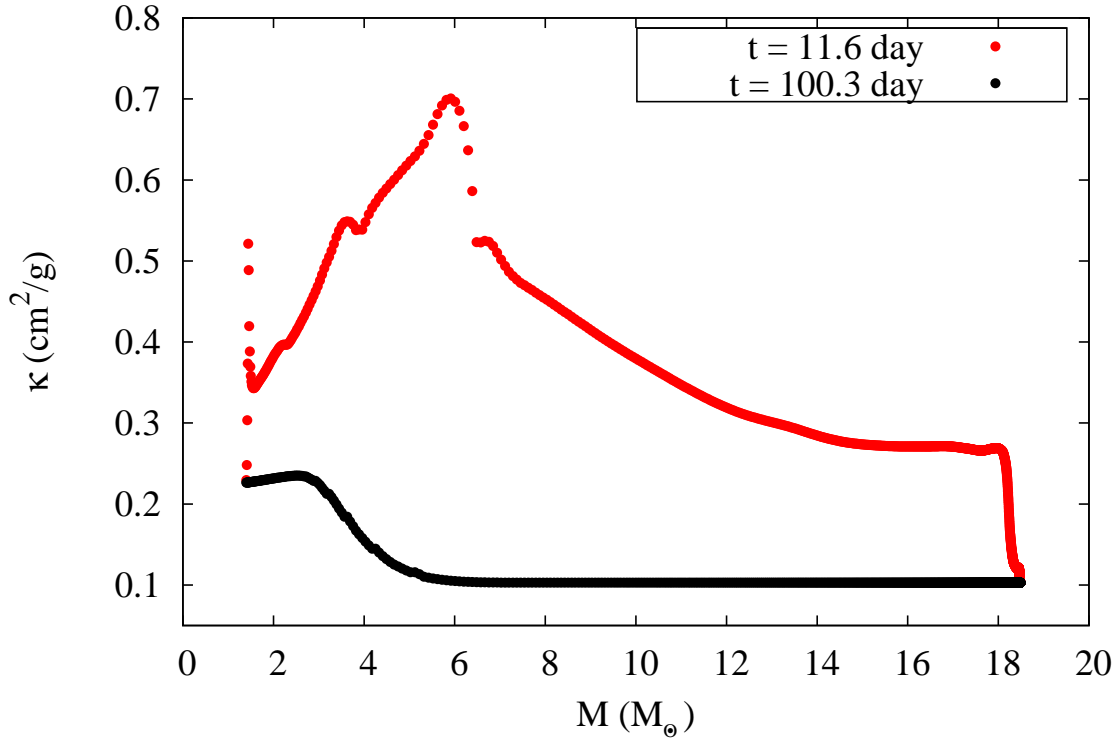}
\caption{The dependence of the Rosseland mean opacity on the mass coordinate of the ejecta at 11.6 day (red) and 100.3 day (black).}
\label{fig:opac}
\end{figure}

\begin{figure}[!ht]
\centering
\plotone{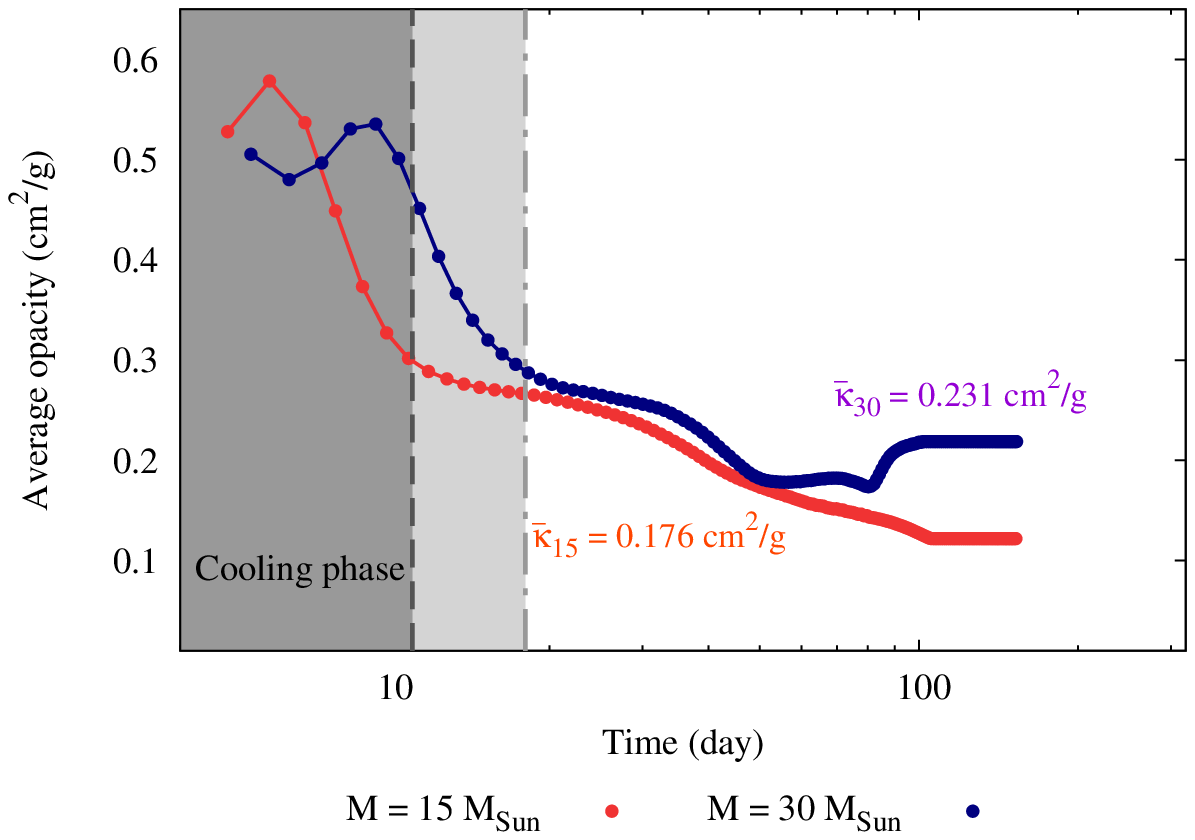}
\caption{The temporal variation of $\kappa(M_{ph})$ for a $15 M_\odot$ (red) and a $30 M_\odot$ (blue) SNEC model. Vertical gray lines represent the time boundaries of the the cooling phases \citep{nagy} for both masses, respectively. The numbers in color indicate the average opacities of different models.}
\label{fig:av}
\end{figure}

In each case the original SNEC opacity output files are used, where the opacity changes both in mass coordinates and time (Fig.~\ref{fig:opac}). In every time-step we integrate the opacity \citep{nagy} from the mass coordinate of the neutron star ($M_0 = 1.34$ M$_\odot$) up to the mass coordinate of the photosphere (M$_{ph}$):
\begin{equation}
\kappa(M_{ph}) = \frac{1}{M_{ph} - M_{0}} \int\limits_{M_0}^{M_{ph}} \kappa\ dm\ .
\label{eq:op}
\end{equation} 
This way we get rid of the space-dependence of the calculated opacities, and receive $\kappa(M_{ph})$ values for all SNEC models that have only temporal variation (Fig.~\ref{fig:av}). 
 
Since the opacity in semi-analytic mo\-dels are constant in space and time, the time-averaged opacity ($\overline{\kappa}$) is defined by integrating the $\kappa(M_{ph})$  values from several days after the shock breakout ($t_0$) up to $t_{end}$ as
\begin{equation}
\overline{\kappa} = \frac{1}{t_{end} - t_0} \int\limits_{t_0}^{t_{end}} \kappa(M_{ph})\ dt\ .
\end{equation}

During our study for Type IIP and Type IIb supernova models we use the so-called two-component configuration, which contains a dense core and an extended, low-mass outer shell \citep[e.g.][]{bersten1,nagy}. Here the first peak of the light curve is caused by the adiabatic cooling of the shock-heated envelope, while the second peak is mainly determined by the radioactive decay of $^{56}$Ni and the recombination of H or He. Thus, we separately calculate the average opacity for both the early cooling phase($\kappa_{shell}$) and the late-time photospheric phase ($\kappa_{core}$). In the early phase, $t_0$ is chosen to be 5 days after the moment of the shock breakout to be sure that the homologous expansion criteria fulfilled, while $t_{end}$ is defined as the termination of the cooling phase when the opacity drops rapidly (gray regions on Fig.~\ref{fig:av}). For the second LC phase, $t_0$ is equal to $t_{end}$ of the early phase, and the integration continues up to the end of the photospheric phase. Moreover to receive comparable result with other semi-analytic models \citep[e.g.,][]{arnett}, we also determine the average opacity by integrating $\kappa(M_{ph})$ from 5 days up to the end of the photospheric phase ($\kappa_{total}$).

\section{Results} \label{sec:result}
\subsection{Type IIP supernova models} \label{sec:iip}
To estimate the average opacity for diffe\-rent core-collapse supernovae we systematically change various physical parameters (initial mass, metallicity, rotation) that determine the mass-loss history of the model star. For simplicity, all stellar models are chosen to create Type IIP-like light curves. Here, we generate progenitor models with an extended mass range (15 - 45 $M_\odot$), because some theoretical models suggest \citep[e.g.][]{heger, nomoto} that upper mass limit for Type IIP-like SNe could be between 40 - 50 $M_\odot$, if these objects go through nickel fallback during black hole formation.

One of the most important parameters that induces changes in the chemical composition and also affects mass-loss, is the initial mass of the progenitor. 
In this case we examine two different approaches: one without any stellar wind processes except mass-loss due to luminosity reaching the Eddington-limit and the other one with MESA 'Dutch' wind-scheme.

\begin{table}[!htb]
\caption{Average opacities for SNEC models with different initial masses (no stellar wind-scheme)} 
\label{table:op1}     
\centering                  
\begin{tabular}{l c c c c }          
\hline
\hline
\noalign{\smallskip}                     
Mass &  $t_{shell}$ ($\mathrm{day}$) & $\overline{\kappa}_{shell}$ ($\mathrm{cm^2/g}$) & $\overline{\kappa}_{core}$ ($\mathrm{cm^2/g}$) & $\overline{\kappa}_{total}$ ($\mathrm{cm^2/g}$)   \\
\noalign{\smallskip}
\hline 
\noalign{\smallskip}
15 $M_\odot$& 11 $\pm$ 1 & 0.364 $\pm$ 0.07 & 0.166 $\pm$ 0.01 & 0.182 $\pm$ 0.01\\ 
20 $M_\odot$& 13 $\pm$ 1 & 0.379 $\pm$ 0.10 & 0.179 $\pm$ 0.01 & 0.199 $\pm$ 0.01\\
25 $M_\odot$& 15 $\pm$ 1 & 0.367 $\pm$ 0.06 & 0.190 $\pm$ 0.03 & 0.210 $\pm$ 0.05\\
30 $M_\odot$& 17 $\pm$ 1 & 0.400 $\pm$ 0.08 & 0.274 $\pm$ 0.05 & 0.295 $\pm$ 0.01\\
35 $M_\odot$& 19 $\pm$ 2 & 0.430 $\pm$ 0.06 & 0.247 $\pm$ 0.05 & 0.274 $\pm$ 0.01\\
40 $M_\odot$& 32 $\pm$ 4 & 0.299 $\pm$ 0.12 & 0.214 $\pm$ 0.07 & 0.240 $\pm$ 0.03\\
45 $M_\odot$& 27 $\pm$ 2 & 0.373 $\pm$ 0.07 & 0.209 $\pm$ 0.06 & 0.232 $\pm$ 0.02\\
\hline                                            
\end{tabular}
\end{table}

\begin{table*}[!ht]
\caption{Average opacities for SNEC models with different initial masses} 
\label{table:op2}     
\centering                  
\begin{tabular}{l c c c c }          
\hline
\hline
\noalign{\smallskip}                     
Mass &  $t_{shell}$ ($\mathrm{day}$) & $\overline{\kappa}_{shell}$ ($\mathrm{cm^2/g}$) & $\overline{\kappa}_{core}$ ($\mathrm{cm^2/g}$) & $\overline{\kappa}_{total}$ ($\mathrm{cm^2/g}$)   \\
\noalign{\smallskip}
\hline 
\noalign{\smallskip}
15 $M_\odot$& 11 $\pm$ 1 & 0.356 $\pm$ 0.07 & 0.158 $\pm$ 0.01 & 0.175 $\pm$ 0.01 \\ 
20 $M_\odot$& 13 $\pm$ 1 & 0.368 $\pm$ 0.07 & 0.172 $\pm$ 0.02 & 0.191 $\pm$ 0.01 \\
25 $M_\odot$& 15 $\pm$ 1 & 0.379 $\pm$ 0.07 & 0.187 $\pm$ 0.02 & 0.208 $\pm$ 0.01 \\
30 $M_\odot$& 15 $\pm$ 1 & 0.352 $\pm$ 0.06 & 0.255 $\pm$ 0.05 & 0.267 $\pm$ 0.01 \\
35 $M_\odot$& 16 $\pm$ 2 & 0.332 $\pm$ 0.03 & 0.218 $\pm$ 0.05 & 0.230 $\pm$ 0.01 \\
40 $M_\odot$& 24 $\pm$ 2 & 0.375 $\pm$ 0.05 & 0.207 $\pm$ 0.06 & 0.237 $\pm$ 0.01 \\
45 $M_\odot$& 17 $\pm$ 2 & 0.365 $\pm$ 0.05 & 0.186 $\pm$ 0.04 & 0.202 $\pm$ 0.01 \\
\hline                                            
\end{tabular}
\end{table*}

Models without any wind-scheme represent extremely low mass-loss rate, but their light curves show typical Type IIP-like structures. Still, in one of the two-component models the derived average opacity is slightly higher than the 0.4 cm$^2$/g theoretical opacity limit for Thompson-scattering (Table~\ref{table:op1}), which means that in this particular configuration we are not able to receive self-consistent average opacities. Apart from this case, the gained opacities show good agreement  with the average chemical composition of the progenitor stars. So, if such a LC fit shows an opacity of  $\kappa \sim$ 0.4 cm$^2$/g, then it is plausible that the progenitor suffered only moderate mass-loss.  

\begin{figure}[!ht]
\centering
\plotone{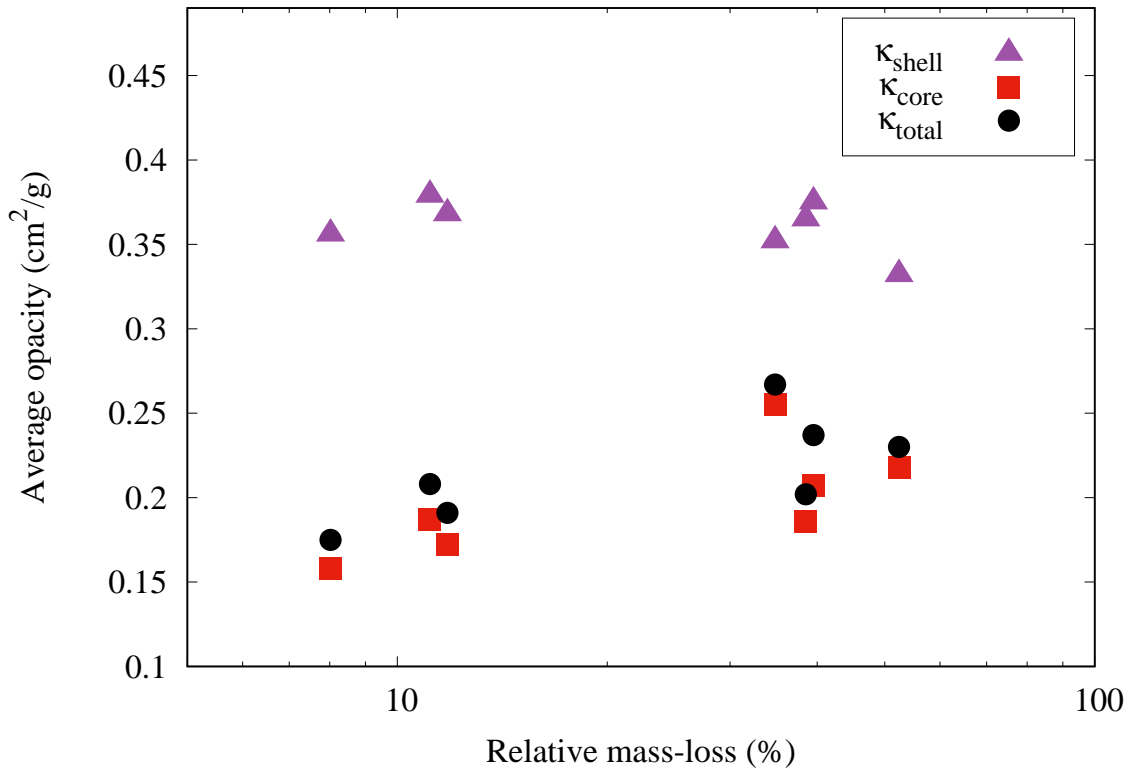}
\caption{ The dependence of the average opacity on the relative mass-loss caused by different initial masses ("Dutch" wind-scheme). The various symbols represent the average opacities from different model approximations: one-component model (circle), shell (triangle) and core (square) configuration.}
\label{fig:op}
\end{figure}

In the second scenario the 'Dutch' wind-scheme is applied to model the mass-loss in the RGB and AGB phases. In MESA this scenario combines the results form \cite{glebbeek}, \cite{vink} and \cite{nugis} to approximate an acceptable mass-loss history for a massive star. If we use this approach the previously mentioned problem can be solved, and we get reasonably good average opacities. As Fig.~\ref{fig:op} shows, models with moderate mass-loss predict lower opacities for both two-component ($\kappa_{shell}$ and $\kappa_{core}$) and single component ($\kappa_{total}$) configuration. As expected, these smaller mass-loss rates represent models with lower masses (Table~\ref{table:op2}), which means that if we want to be self-consistent while modeling the supernova LCs, we have to use slightly lower opa\-cities (0.18 - 0.2 cm$^2$/g) than usual for Type IIP SNe with 8 - 10 M$_\odot$ ejecta.  

The intensity of the stellar wind could also be an important parameter. In MESA we are able to change the strength of the wind-scheme with a scaling factor ($\eta$) as
\begin{equation}
\dot{M} = \eta\ \dot{M}_{Dutch}\ ,
\label{eq:m}
\end{equation}
where $\dot{M_{Dutch}}$ and $\dot{M}$ is the originally calculated and the finally used mass-loss rate, respectively. 
Nevertheless, the results of this analysis show (Table~\ref{table:op3}) that the stellar wind intensity influence neither the generated LCs nor the calculated average opa\-cities significantly. So, the intensity of the mass-loss processes cannot be determined by fitting the LC of the SNe.

\begin{table}[!htb]
\caption{Average opacities for SNEC models with different wind-scheme scaling factors} 
\label{table:op3}     
\centering                  
\begin{tabular}{l c c c c }          
\hline
\hline
\noalign{\smallskip}                     
$\eta$ & $t_{shell}$ ($\mathrm{day}$) & $\overline{\kappa}_{shell}$ ($\mathrm{cm^2/g}$) & $\overline{\kappa}_{core}$ ($\mathrm{cm^2/g}$) & $\overline{\kappa}_{total}$ ($\mathrm{cm^2/g}$)   \\
\noalign{\smallskip}
\hline 
\noalign{\smallskip}
0.2 & 12 $\pm$ 1 & 0.377 $\pm$ 0.11 & 0.182 $\pm$ 0.01 & 0.199 $\pm$ 0.02 \\
0.4 & 12 $\pm$ 1 & 0.377 $\pm$ 0.11 & 0.182 $\pm$ 0.01 & 0.199 $\pm$ 0.02 \\
0.6 & 13 $\pm$ 1 & 0.365 $\pm$ 0.10 & 0.178 $\pm$ 0.02 & 0.195 $\pm$ 0.01 \\
0.8 & 13 $\pm$ 1 & 0.366 $\pm$ 0.09 & 0.175 $\pm$ 0.02 & 0.193 $\pm$ 0.01 \\
1.0 & 14 $\pm$ 2 & 0.355 $\pm$ 0.08 & 0.173 $\pm$ 0.02 & 0.191 $\pm$ 0.01 \\
\hline                                            
\end{tabular}
\end{table}

\begin{table}[!htb]
\caption{Average opacities for SNEC models with different metallicities} 
\label{table:op4}     
\centering                  
\begin{tabular}{l c c c c c }          
\hline
\hline
\noalign{\smallskip}   
Z/$Z_\odot$ &  $t_{shell}$ ($\mathrm{day}$) & $\overline{\kappa}_{shell}$ ($\mathrm{cm^2/g}$) & $\overline{\kappa}_{core}$ ($\mathrm{cm^2/g}$) & $\overline{\kappa}_{total}$ ($\mathrm{cm^2/g}$) \\
\noalign{\smallskip}
\hline 
\noalign{\smallskip}
4.0 & 15 $\pm$ 1 & 0.333 $\pm$ 0.07 & 0.169 $\pm$ 0.02 & 0.191 $\pm$ 0.01\\
1.0 & 13 $\pm$ 1 & 0.368 $\pm$ 0.07 & 0.172 $\pm$ 0.02 & 0.190 $\pm$ 0.01\\
0.1 & 70 $\pm$ 5 & 0.362 $\pm$ 0.10 & 0.152 $\pm$ 0.23 & 0.245 $\pm$ 0.09\\
0.01 & 59 $\pm$ 3 & 0.406 $\pm$ 0.09 & 0.173 $\pm$ 0.17 & 0.245 $\pm$ 0.13\\
0.001 & 60 $\pm$ 3 & 0.396 $\pm$ 0.06 & 0.173 $\pm$ 0.19 & 0.248 $\pm$ 0.13\\
\hline                                            
\end{tabular}
\end{table}

\begin{figure}[!ht]
\centering
\plotone{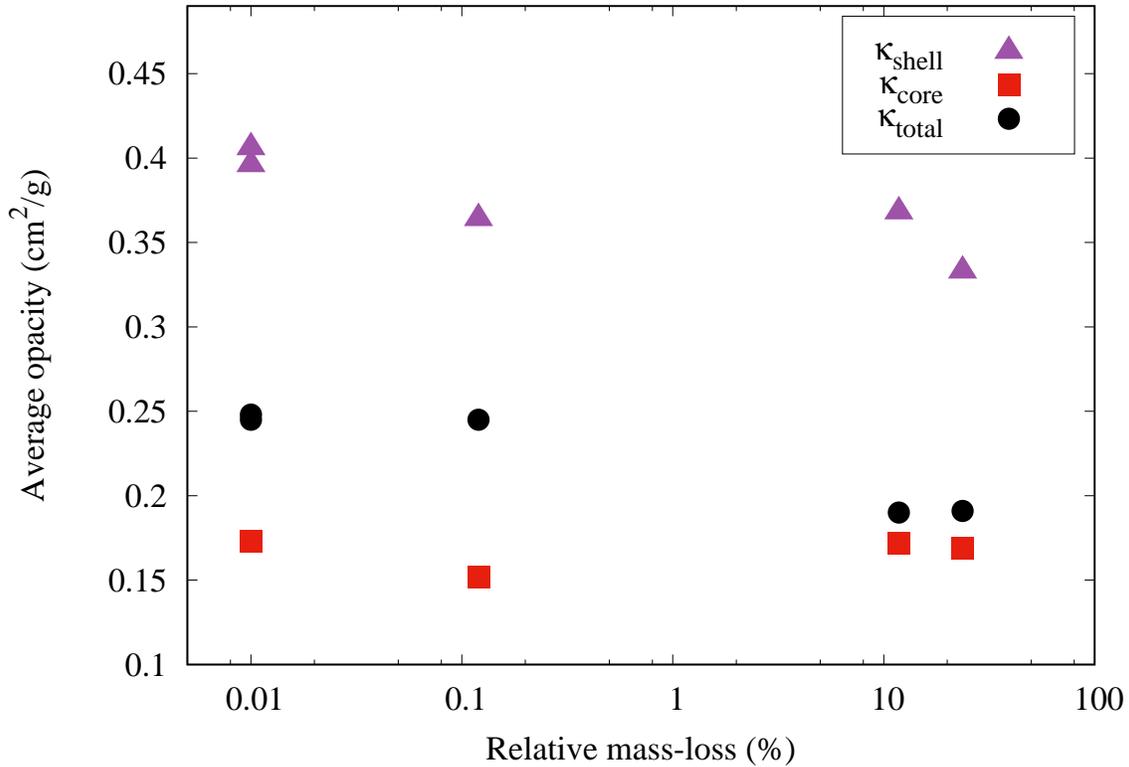}
\caption{ The dependence of the average opacity on the relative mass-loss caused by different metallicities. The various symbols represent the average opacities from different model approximations: one-component model (circle), shell (triangle) and core (square) configuration.}
\label{fig:op1}
\end{figure}

Another significant physical parameter which indicates mass-loss during the stellar evolution, is the metallicity of the exploding star. The results show that, as we expect, stars with lower metallicity are able to keep most of their hydrogen and helium layers (Fig.~\ref{fig:op1}). Thus, the opacity values for low-metallicity stars become higher than opacities of a solar-like object (Table~\ref{table:op4}). So, if during LC fitting we get average opacities above $\kappa \sim$ 0.35 cm$^2$/g, then it is possible that the metal content of the progenitor is somewhat lower than the solar abundance of the heavier elements. However, it should keep in mind that this opacity region is similar to the opacity range from moderate mass-loss, which make the metallicity estimation quite uncertain.          

\begin{table*}[!htb]
\caption{Average opacities for SNEC models with different scaling factors for critical surface velocity} 
\label{table:op5}     
\centering                  
\begin{tabular}{l c c c c }          
\hline
\hline
\noalign{\smallskip}                     
$\zeta$ & $t_{shell}$ ($\mathrm{day}$) & $\overline{\kappa}_{shell}$ ($\mathrm{cm^2/g}$) & $\overline{\kappa}_{core}$ ($\mathrm{cm^2/g}$) & $\overline{\kappa}_{total}$ ($\mathrm{cm^2/g}$)   \\
\noalign{\smallskip}
\hline 
\noalign{\smallskip}
0.0 & 15 $\pm$ 1 & 0.358 $\pm$ 0.06 & 0.169 $\pm$ 0.02 & 0.193 $\pm$ 0.01 \\
0.2 & 15 $\pm$ 1 & 0.375 $\pm$ 0.08 & 0.180 $\pm$ 0.02 & 0.199 $\pm$ 0.01 \\
0.4 & 14 $\pm$ 1 & 0.362 $\pm$ 0.08 & 0.174 $\pm$ 0.02 & 0.193 $\pm$ 0.01 \\
0.6 & 14 $\pm$ 1 & 0.362 $\pm$ 0.09 & 0.174 $\pm$ 0.02 & 0.193 $\pm$ 0.01 \\
0.8 & 13 $\pm$ 1 & 0.366 $\pm$ 0.09 & 0.175 $\pm$ 0.02 & 0.193 $\pm$ 0.01 \\
\hline                                            
\end{tabular}
\end{table*}

The surface rotation of the star may influence mass-loss as well. In MESA we use the break-up velocity ($v_{break}$) as the surface velocity ($v_{surf}$) of the model start, and during the calculations we are able to change the strength of the surface rotation with a scaling factor ($\zeta$) as
\begin{equation}
v_{surf} = \zeta\ v_{break} \approx \zeta\ \sqrt[]{\frac{G\ M}{R}}\ .
\label{eq:v}
\end{equation}
As it can be seen in Table~\ref{table:op5}, the average opacities are not influenced significantly by the intensity of the surface velocity. Thus, from LC modeling the rotation of the progenitor cannot be constrained.

\begin{figure}[!ht]
\centering
\plotone{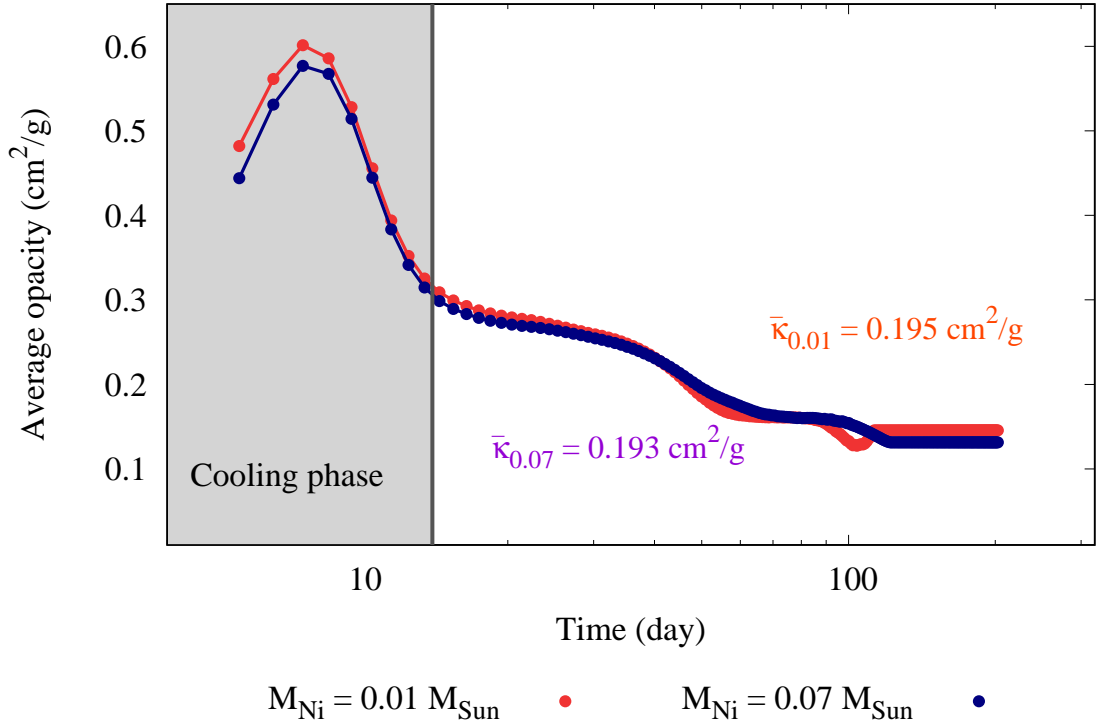}
\caption{The temporal variation of $\kappa(M_{ph})$ for a $M_{Ni} = 0.01 M_\odot$ (red) and a $M_{Ni} = 0.07 M_\odot$ (blue) SNEC model. Vertical gray lines represent the time boundaries of the the cooling phases \citep{nagy} for both masses, respectively. The colored numbers indicate the different average opacities of the two models.}
\label{fig:ni}
\end{figure}

The initial nickel mass could be an important parameter as well, because it changes the ratio of the heavy elements within the ejecta, which may cause the average opacity to decrease. Nevertheless, our results show (Fig.~\ref{fig:ni}) that the initial nickel mass only slightly influences the $\kappa_{M_{ph}}$ values, which means that after integration the received average opa\-cities are basically the same within error bars. Although the light curves of the SNe are significantly affected by $M_{Ni}$, the average opacity of these events do not depend significantly on the initial nickel mass of the ejecta.

\subsection{Stripped-envelope SNe}\label{sec:sesn}
All previous results refer only to Type IIP-like progenitors. For stripped-envelope SNe (Type IIb, Ib and Ic), however, the mass-loss mechanisms could be quite different, because it is plausible that the progenitors of these explosions may have a binary companion \citep[e.g.,][]{moriya,yoon}. For example, according to study of \cite{yoon} the outer H layers of Type IIb SN progenitors are more likely removed by Roche-lobe overflow rather than by stellar wind. Thus, to determine the average opacities for stripped-envelope SNe, we need to take into consideration the extreme mass-loss via binary interaction. 

First, to compare the average opacity values for diverse types of stripped-envelope SNe, we just mimic the global stellar structure of their progenitors.
In order to estimate the progenitor of a Type IIb, Ib and Ic SN, we simply remove the outer envelope of a MESA model star manually. For the Type IIb model most of the outer H-rich envelope is cut off, so only $\sim 1$ M$_{\odot}$ of hydrogen remained. For the Type Ib model we remove the total H layer of the star, while for the Type Ic model we detach both the H and He envelopes. 

Second, we create a more self-consistent model for a Type IIb explosion, which contains an interacting binary system. Here we use the binary module of MESA, where the first component is the progenitor star with $20 M_\odot$ initial mass, and the second component is the acceptor star ($M = 10 M_\odot$). Nevertheless, to reduce numerical and run-time errors we only calculate the post-RGB phase of the evolution of this binary system, which means that we model the early evolution of both stars as individual objects.    

Because our main goal is to explore the effect of binary interaction, we choose only the default prescriptions in MESA: Schwarzschild convection, "Dutch" wind-scheme with $\eta = 0.8$ and "approx21" nuclear reaction network, but we do not include any magnetic effects. The initial rotational period of the binary is 50 day, which decreases rapidly during the stellar evolution. During the interacting phase we use the so-called "Kolb" mass-loss scheme that describe an optically thick overflow into the binary system \citep{kolb}. 

It can be seen in Table~\ref{tab:typ} that during the cooling phase, $\overline{\kappa}$ is 0.4 cm$^2$/g for a typical Type IIP SN with a massive H-rich ejecta. However, the average opacity decreases to around 0.3 cm$^2$/g for both Type IIb models, which corresponds to a star that lost most of its H-rich envelope. In contrast, during the later phase, the average opacity of Type IIP and IIb supernovae is considerably similar, having a value of about 0.2 cm$^2$/g. 
Comparing the average opacities from the two different Type IIb models as well, it can be seen that $\overline{\kappa}$ is approximately equal in both cases. An acceptable explanation of this result could be that the binary overflow do not change the global chemical structure of the donor star, it just rips off its outer layers. So, at the time of the core-collapse the chemical abundances of such an object are approximately similar to the inner regions of a single massive star. Therefore, it could be a plausible simplification to use the average opacities from the cut-off models for fitting the light curves of stripped-envelope SN light curves.

Although the two-component configuration is not a adequate solution for Type Ib and Ic events, the gained $\overline{\kappa}_{total}$ values can be comparable with the average opacities from Type IIP and IIb model calculations. For Type Ib the average opacities are only slightly lower, while in Type Ic SNe these are a factor of two lower, which agree well to the mass-loss history of these objects (Table~\ref{tab:typ}).

\begin{deluxetable*}{lccccc}[!htb]
\tablecaption{Average opacities for different types of CCSNe}                 
\tablehead{                     
Parameter & IIP & IIb\tablenotemark{1} & IIb\tablenotemark{2} & Ib & Ic }
\startdata
$M_{ej}$ ($\mathrm{M_\odot}$)& 16.5 & 5.9 & 6.01& 3.0& 2.0\\
$t_{shell}$ ($\mathrm{day}$) & 13 $\pm$ 1 & 10 $\pm$ 1 & 9 $\pm$ 1 & - & -\\ 
$\overline{\kappa}_{shell}$ ($\mathrm{cm^2/g}$)& 0.381 $\pm$ 0.01 & 0.298 $\pm$ 0.02 & 0.293 $\pm$ 0.02 & - & - \\
$\overline{\kappa}_{core}$ ($\mathrm{cm^2/g}$)& 0.20 $\pm$ 0.01 & 0.194 $\pm$ 0.01 & 0.193 $\pm$ 0.01 & 0.182 $\pm$ 0.01 & 0.10 $\pm$ 0.01 \\
$\overline{\kappa}_{total}$ ($\mathrm{cm^2/g}$)& 0.213 $\pm$ 0.03 & 0.195 $\pm$ 0.01 & 0.195 $\pm$ 0.02 & 0.182 $\pm$ 0.01 & 0.10 $\pm$ 0.01 \\
\enddata
\tablenotetext{1}{Binary model for Type IIb SN}\tablenotetext{2}{Cut off model for Type IIb SN}
\label{tab:typ}
\end{deluxetable*}

\section{Verification}
\begin{figure}[!ht]
\centering
\plotone{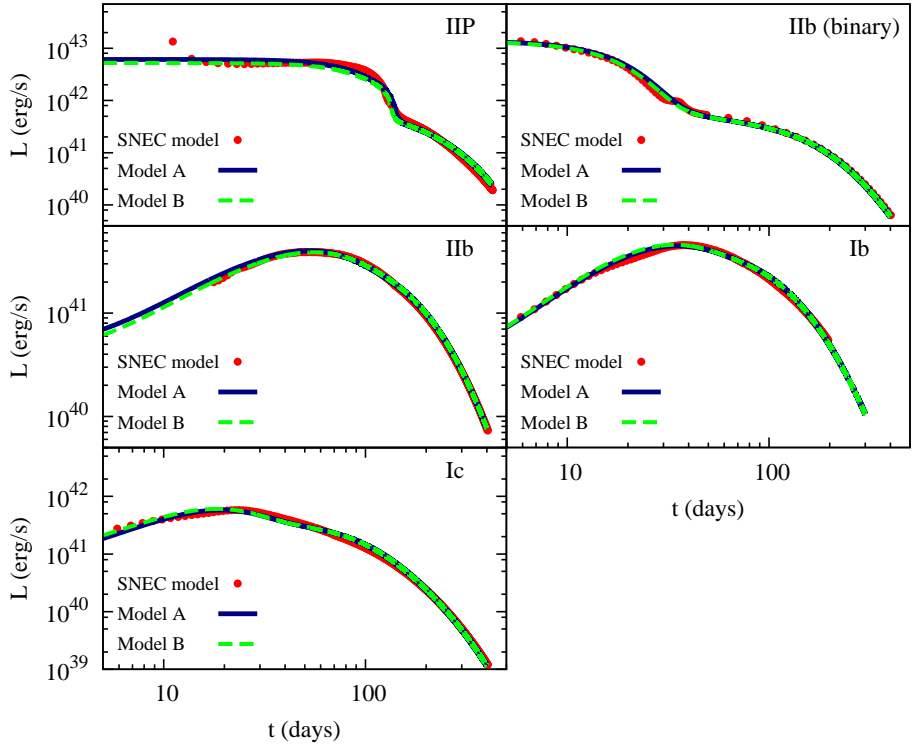}
\caption{Bolometric light curve fits of different CCSNe from SNEC (red). Model A (blue) and B (green) represent the calculation with fixed $\overline{\kappa}_{total}$ and $\kappa_{Th}$, respectively (Table~\ref{tab:comp}).}
\label{fig:lc}
\end{figure}

As we show in Sec.~\ref{sec:sesn}, the average opacities calculated from SNEC models are in the ballpark of the generally used Thompson-scattering opacities in the literature ($\kappa_{Th}$). To explore the applicability of these gained opacities we fit SNEC model LCs with our semi-analytic code \citep{nagy} using fixed $\overline{\kappa}_{total}$ (Model A) and $\kappa_{Th}$ (Model B) values to get the same fit-by-eye synthetic LCs (Fig.~\ref{fig:lc}), and compare the fitting parameters with the initial hydrodynamic properties, namely kinetic energy and ejected mass (Table~\ref{tab:comp}). For more consistent comparison, we only change $E_{kin}$ and $M_{ej}$ in the models, while the other fitting parameters (e.g. initial nickel mass, initial radius of the progenitor) are the same. Because of the parameter correlation, the relative deviations of these individual parameters do not show any significant differences in the various models. To reduce the correlation between $E_{kin}$ and $M_{ej}$, we examine the combination of them by introducing the so-called scaling velocity \citep{arnett}, which is 
\begin{equation}
v_{sc} = \sqrt{\frac{10\ E_{kin}}{3\ M_{ej}}}
\end{equation}
for an ejecta with constant density profile. For comparison, this scaling velocity is also calculated for the different SNEC models. Although in hydrodynamic calculations the density profile of the ejecta is not constant at all, its variation is quite stale and the calculated velocities show reasonably good agreement with the typical expansion velocities of CCSNe. This approximation seems to be acceptable for further study. The only exception is the Type IIb binary model, where $v_{sc}$ is approximately an order-of-magnitude lower than the expected one. Thus, in this case we use the maximum photospheric velocity for the comparison, which is roughly equal to the scaling velocity if the density profile is steep. 

\begin{deluxetable}{llccccccc}[!htb]
\tablecaption{Parameter comparison for different types of CCSNe}                 
\tablehead{ 
\noalign{\smallskip}
 & & $\kappa$ ($\mathrm{cm^2/g}$) & $M_{ej}$ ($\mathrm{M_\odot}$) & $\Delta M_{ej}$ ($\mathrm{\%}$)& $E_{kin}$ ($\mathrm{10^{51} erg}$) & $\Delta E_{kin}$ ($\mathrm{\%}$)& $v_{sc}$ ($\mathrm{km/s}$) & $\Delta v_{sc}$ ($\mathrm{\%}$)}
\startdata
& SNEC & - & 16.5 & - & 3.1 & - & 5611 & - \\
IIP & Model A & 0.213 & 7.5 & 54.4 & 1.6 & 48.4 & 5979 & 6.6 \\ 
& Model B & 0.34 & 5.5 & 66.7 & 1.6 & 48.4  & 6982 & 24.4 \\
\hline
& SNEC & - & 5.9 & - & 2.02 & - & 11725 & - \\
IIb\tablenotemark{1} & Model A & 0.195 & 1.45 & 75.4 & 1.2 & 40.6 & 11776 & 0.4 \\ 
& Model B & 0.24 & 1.2 & 79.7 & 1.2 & 40.6 & 12945 & 10.4 \\
\hline
& SNEC & - & 6.01 & - & 2.06 & - & 7573 & - \\
IIb\tablenotemark{2} & Model A & 0.195 & 6.0 & 0.17 & 2.0 & 2.9 & 7474 & 1.3 \\
& Model B  & 0.24 & 5.9 & 1.8 & 2.2 & 6.8 & 7905 & 4.4 \\
\hline
& SNEC & - & 3.0 & - & 1.01 & - & 7511 & - \\
Ib & Model A & 0.182 & 3.2 & 6.6 & 1.0 & 0.99 & 7236 & 3.7 \\ 
& Model B & 0.2 & 3.3 & 10 & 1.5 & 48.5 & 8728 & 16.2 \\
\hline
& SNEC & - & 2.0 & - & 1.02 & - & 9245 & - \\
Ic & Model A & 0.1 & 2.2 & 10 & 1.1 & 7.8 & 9154 & 0.98 \\ 
& Model B & 0.06 & 2.9 & 45 & 1.1 & 7.8 & 7973 & 13.8 \\
\enddata
\tablenotetext{1}{Binary model for Type IIb SN}\tablenotetext{2}{Cut off model for Type IIb SN}
\label{tab:comp}
\end{deluxetable}

As it can be seen in Table~\ref{tab:comp}, the scaling velocities for the models with $\kappa_{Th}$ display considerably higher relative deviation ($\Delta v_{sc}$) than the fits with $\overline{\kappa}_{total}$. Thus, the new average opacities calculated in this paper could improve the parameter estimation using semi-analytic model LC fits to CCSNe. 

\section{Conclusions}
Although the constant Thompson-scattering opacity is not a perfect estimation for core-collapse supernova explosions because of the rapidly changing opacities in their ejecta, the calculated average opacities show reasonably good agreement with frequently used constant opacities in the literature \citep{nakar,huang}. Moreover, our results indicate that the two-component confi\-guration could be re\-levant for modeling Type IIb and IIP SNe, because the derived average opacities for both the shell and the core component are similar to the expected values from the average che\-mical composition.  

On the other hand, if we choose the opacity wisely during model fitting, we may estimate roughly the chemical composition of the progenitor. But it should keep in mind that, because of the correlation of the model parameters, we are not able to recover the exact confi\-guration of the exploding star from only the applied opacity va\-lues.

\acknowledgments
This research is supported by the GINOP-2.3.2-15-2016-00033 project which is funded by the Hungarian National Research, Development and Innovation Fund together with the European Union.

\software{MESA \citep{paxton11,paxton,paxton15,paxton18}, SNEC \citep{morozova}, LC2 \citep{nagy}}



\end{document}